\documentclass[preprint,showpacs,preprintnumbers,amsmath,amssymb,showkeys]{revtex4}

\usepackage{dcolumn}% Align table columns on decimal point
\usepackage{bm}% bold math
\usepackage{amssymb,amsmath}

\begin{document}

%\preprint{APS/123-QED}

\title{Dual Kappa Poincare Algebra}

\author{Jose A. Magpantay}
\email{jose.magpantay@up.edu.ph} \affiliation{National Institute
of Physics and Technology Management Center, University of the Philippines, Quezon City,
Philippines\\}

\date{\today}

\begin{abstract}
We show a different modification of Poincare algebra that also preserves the Lorentz algebra. The change begins with  how boosts affect space-time in a way similar to how boosts affect the momenta in kappa Poincare algebra, thus the name dual kappa Poincare algebra. Since by construction the new space-time commutes, it follows that the momenta co-commute. Proposing a space-time co-algebra that is similar to the momentum co-product in the bicrossproduct basis of kappa Poincare algebra, the phase space algebra is derived using the Heisenberg double construction. The phase space variables of the dual kappa Poincare algebra are then related to SR phase space variables. From these relations, we complete the dual kappa Poincare algebra by deriving the action of rotations and boosts on the momenta.      
\end{abstract}

\maketitle

\section{\label{sec:level1}The Poincare Algebra}
	The generators of the Poincare algebra are $m_i$, $n_i$ and $p_\mu$, which correspond to real rotations, boosts and translations. The algebra is given by the relations
\begin{subequations}\label{sha}
\begin{align}
\left[m_i,m_j\right]=i\epsilon_{ijk}m_k,\label{first}\\
\left[m_i,n_j\right]=i\epsilon_{ijk}n_k,\label{second}\\
\left[n_i,n_j\right]=-i\epsilon_{ijk}m_k,\label{third}\\
\left[m_i,p_0\right]=0,\label{fourth}\\
\left[m_i,p_j\right]=i\epsilon_{ijk}p_k,\label{fifth}\\
\left[n_i,p_0\right]=ip_i,\label{sixth}\\
\left[n_i,p_j\right]=i\delta_{ij}p_0
\end{align}
\end{subequations}
These algebraic relations follow if we consider the basic phase space brackets
\begin{subequations}\label{sha2}
\begin{align}
\left[x_\mu,p_\nu\right]=i\eta_{\mu\nu},\label{first}\\
\left[x_\mu,x_\nu\right]=\left[p_\mu,p_\nu\right]=0,\label{second}\\
\eta_{\mu\nu}=\left(-,+,+,+\right),
\end{align}
\end{subequations}
and the defining equations for rotations and boosts
\begin{subequations}\label{sha3}
\begin{align}
m_i=\epsilon_{ijk}x_jp_k,\label{first}\\
n_i=x_ip_0-x_0p_i.
\end{align}
\end{subequations}
From \eqref{sha2} and \eqref{sha3}, we can derive the action of rotations and boosts on space-time
\begin{subequations}\label{sha4}
\begin{align}
\left[m_i,x_0\right]=0,\label{first}\\
\left[m_i,x_j\right]=i\epsilon_{ijk}x_k,\label{second}\\
\left[n_i,x_0\right]=ix_i,\label{third}\\
\left[n_i,x_j\right]=i\delta_{ij}x_0.
\end{align}
\end{subequations}
These equations show that the coordinates $x_i$ and momenta $p_i$ rotate while $x_0$ (time) and $p_0$ (energy) are invariant under real rotations. Note also that the boosts act on the momenta and and coordinates in the same way.

\section{\label{sec:level2}The Kappa Poincare Algebra}
	The deformation of the Poincare algebra has evolved from the pioneering work of Lukierski and collaborators  \cite{Lukierski1} to the present form \cite{Majid} where the rotation generators $M_i$, the boosts $N_i$ and the momenta $P_\mu$ satisfy a deformed Poincare algebra measured by a deformation parameter $\kappa$ such that (i) the Lorentz algebra is maintained (the first three relations of \eqref{sha}), (ii) the rotation generators' action on the momenta remains the same (the fourth and fifth relations in \eqref{sha}), and (iii) the boosts action on momentum (the last two equations of \eqref{sha}) is changed. The first is easily implemented by    
\begin{subequations}\label{sha5}
\begin{align}
M_i=m_i,\label{first}\\
N_i=n_i.
\end{align}
\end{subequations}
The second is implemented by 
\begin{subequations}\label{sha6}
\begin{align}
\left[M_i,P_0\right]=0,\label{first}\\
\left[M_i,P_j\right]=i\epsilon_{ijk}P_k.
\end{align}
\end{subequations}
Using \eqref{sha5} and the SR commutators, we see that \eqref{sha6} is satisfied if
\begin{subequations}\label{sha7}
\begin{align}
P_0=f(p_0,\vec{p}\cdot\vec{p}),\label{first}\\
P_i=p_ig(p_0,\vec{p}\cdot\vec{p}).
\end{align}
\end{subequations}
As for the action of boosts on the momenta, these are given by
\begin{subequations}\label{sha8}
\begin{align}
\left[N_i,P_0\right]=iP_iD(P_0,\vec{P}\cdot\vec{P}),\label{first}\\
\left[N_i,P_j\right]=i\delta_{ij}A(P_0,\vec{P}\cdot\vec{P})+iP_iP_jB(P_0,\vec{P}\cdot\vec{P}).
\end{align}
\end{subequations}
By tensor structure, these are the only possible terms along with an $\epsilon_{ijk}P_k$, which can be shown to vanish via Jacobi's identity. The functions A, B and D are dependent on a deformation parameter $\kappa$ and must get back the Poincare algebra in a suitable limit (which we will show later as $\kappa$ approaches $\infty$). This means B approaches zero, D approaches one and A approaches $P_0$ in the Poincare limit.  

It is usually argued that the deformation of the Poincare algebra is not unique because the three functions A, B and D can be shown to satisfy the non-linear equation \cite{Kowalski-Glikman1}
\begin{equation}\label{sha9}
\frac{\partial{A}}{\partial{P_0}}D+2\frac{\partial{A}}{\partial{(\vec{P}\cdot\vec{P})}}\left[A+(\vec{P}\cdot\vec{P})B\right]-AB=1,
\end{equation}
which leaves two functions out of three that is not specified. Actually, the reason is more restrictive, there is only one function that is not specified and it is not A, B or D but either f or g of \eqref{sha7}. Making use of the second of equation \eqref{sha5}, the fundamental commutators in SR phase space, equation \eqref{sha7} and its inverse as given by
\begin{subequations}\label{sha10}
\begin{align}
p_0=F(P_0,\vec{P}\cdot\vec{P}),\label{first}\\
p_i=P_iG(P_0,\vec{P}\cdot\vec{P}),
\end{align}
\end{subequations}
we find that
\begin{subequations}\label{sha11}
\begin{align}
A(P_0,\vec{P}\cdot\vec{P})= p_0g(p_0,\vec{p}\cdot\vec{p}),\label{first}\\
B(P_0,\vec{P}\cdot\vec{P})=G^2\left[2F\frac{\partial{g}}{\partial{(\vec{p}\cdot\vec{p})}}+\frac{\partial{g}}{\partial{p_0}}\right],\label{second}\\
D(P_0,\vec{P}\cdot\vec{P})=G\left[2F\frac{\partial{f}}{\partial{(\vec{p}\cdot\vec{p})}}+\frac{\partial{f}}{\partial{p_0}}\right].
\end{align}
\end{subequations}
At the right hand side of equation \eqref{sha11}, we have to substitute the expressions given by \eqref{sha10} to give the $P_0$ and $\vec{P}\cdot\vec{P}$ dependence of A, B and D. Presenting the deformation of Poincare algebra in terms of relations with SR phase space variables, we find that of the two functions f and g (or the inverses F and G), only one function is not specified because of equation \eqref{sha9}. Equation \eqref{sha11} has not been given in the literature as far as the author knows. Since there is one unspecified function, there is no unique deformation of the Poincare algebra and there are four that that are currently cited in the literature - the bicrossproduct basis or DSR1 \cite{Majid}, the Magueijo-Smolin or DSR2 \cite{Magueijo}, the classical basis \cite{Maslanka} and the earliest, which is the Snyder basis \cite{Snyder}.

	At this stage, it looks like DSR is merely SR in a non-linear momentum representation. The complete dissociation from SR and Poincare algebra is made when a non-trivial co-algebra is endowed on the momentum sector. The fact that the momentum sector commutes by construction (see \eqref{sha7}) results in a space-time sector that co-commutes, i.e.,
\begin{equation}\label{sha12}
\Delta(X_\mu)=\textbf{1}\otimes X_\mu+X_\mu\otimes\textbf{1}.
\end{equation}
Endowing the momentum sector with a non-trivial co-algebra as in the bicrossproduct basis or DSR1   
\begin{subequations}\label{sha13}
\begin{align}
\Delta(P_0)=P_0\otimes\textbf{1}+\textbf{1}\otimes P_0,\label{first}\\
\Delta(P_i)=P_i\otimes\textbf{1}+\exp{(-\frac{P_0}{\kappa})}\otimes P_i,
\end{align}
\end{subequations} 
and making use of the Heisenberg double \cite{Lukierski2}, we find the phase space algebra
\begin{subequations}\label{sha14}
\begin{align}
\left[X_0,P_0\right]=-i,\label{first}\\
\left[X_i,P_j\right]=i\delta_{ij},\label{second}\\
\left[X_i,X_j\right]=\left[P_0,X_i\right]=0,\label{third}\\
\left[X_0,P_i\right]=\frac{i}{\kappa}P_i,\label{fourth}\\
\left[X_0,X_i\right]=-\frac{i}{\kappa}X_i.
\end{align}
\end{subequations}
The first four relations of \eqref{sha14} are only valid for DSR1 while the last seems to be universal for all DSR theories \cite{Kowalski-Glikman1}. Equation (14e) shows the Lie algebraic, non-commuting nature of DSR space-time. This result along with the proof of Kowalski-Glikman \cite{Kowalski-Glikman2} that the momentum space of DSR is De Sitter completes the proof that DSR describes physics beyond SR.

To complete the bicrossproduct Poincare algebra, we will derive the action of rotations and boosts on space-time making use of the DSR1 $\leftrightarrow$ SR phase space transformations. The explicit f and g or F and G are
\begin{subequations}\label{sha15}
\begin{align}
f=\kappa\ln{(\frac{p_0}{\kappa}+\sqrt{1+\frac{m^2}{\kappa^2}})},\label{first}\\
g=\dfrac{1}{\frac{p_0}{\kappa}+\sqrt{1+\frac{m^2}{\kappa^2}}},\label{second}\\
F=\kappa\sinh{\frac{P_0}{\kappa}}+\frac{1}{2\kappa}(\vec{P}\cdot\vec{P})\exp{(\frac{P_0}{\kappa})},\label{third}\\
G=\exp{(\frac{P_0}{\kappa})}.
\end{align}
\end{subequations}
Using \eqref{sha11}, we will get
\begin{subequations}\label{sha16}
\begin{align}
A(P_0,\vec{P}\cdot\vec{P})=\frac{\kappa}{2}\left(1-\exp{(\frac{-2P_0}{\kappa})}\right)+\frac{1}{2\kappa}\vec{P}\cdot\vec{P},\label{first}\\
B(P_0,\vec{P}\cdot\vec{P})=-\frac{1}{\kappa},\label{second}\\
D(P_0,\vec{P}\cdot\vec{P})=1,
\end{align}
\end{subequations}
which we are citing for completeness of the DSR1 relations and to show that in the limit $\kappa \rightarrow \infty$, we get back the Poincare limit. 

The phase space algebra given by \eqref{sha14} is satisfied by the following expression for the DSR1 space-time in terms of SR phase space variables
\begin{equation}\label{sha17}
X_\mu=x_\mu\left(\frac{p_0}{\kappa}+\sqrt{1+\frac{m^2}{\kappa^2}}\right).
\end{equation}
Three comments are in order at this point. First, a similar relation was given in the context of DSR2 \cite{Ghosh}. Second, the transformation is not purely space-time, it involves the SR momenta to give space-time non-commutativity. Third, the DSR $\leftrightarrow $ SR transformation is not canonical because the fundamental brackets are not invariant \cite{Goldstein} again emphasizing that DSR is not SR in a non-linear basis.

Using \eqref{sha17}, \eqref{sha15}, \eqref{sha5} and \eqref{sha3}, we find 
\begin{subequations}\label{sha18}
\begin{align}
M_i=\epsilon_{ijk}X_jP_k,\label{first}\\
N_i=\frac{\kappa}{2}X_i\left(1-\exp{(\frac{-2P_0}{\kappa})}\right)+\frac{1}{2\kappa}X_i(\vec{P}\cdot\vec{P})-X_0P_i.
\end{align}
\end{subequations}
Using these relations and \eqref{sha14}, we find
\begin{subequations}\label{sha19}
\begin{align}
\left[M_i,X_0\right]=0,\label{first}\\
\left[M_i,X_j\right]=i\epsilon_{ijk}X_k,\label{second}\\
\left[N_i,X_0\right]=iX_i-\frac{i}{\kappa}N_i,\label{third}\\
\left[N_i,X_j\right]=i\delta_{ij}X_0-\frac{i}{\kappa}\epsilon_{ijk}M_k.
\end{align}
\end{subequations}
To summarize, the kappa Poincare algebra is given by the Lorentz algebra (guaranteed by \eqref{sha5}) and \eqref{sha6} and \eqref{sha8}. In particular, the bicrossproduct basis is given by \eqref{sha15} and \eqref{sha16} with the coalgebra given by \eqref{sha12} and \eqref{sha13}. These results in the phase space algebra given by \eqref{sha14}, from which we deduce \eqref{sha17}. The bicrossproduct relations is completed by deriving \eqref{sha19} from \eqref{sha18} and \eqref{sha14}, which can be deduced directly from the co-algebra of rotations and boosts \cite{Kowalski-Glikman1}. 

\section{\label{sec:level3}The Dual Kappa Poincare Algebra}
	To present the dual kappa Poincare algebra, we begin by defining its generators as $\bar{P_\mu}$, $\bar{M_i}$ and $\bar{N_i}$. Just as in the Poincare and kappa Poincare algebra, we will maintain the Lorentz algebra and we can do this by identifying
\begin{subequations}\label{sha20}
\begin{align}
\bar{M_i}=m_i,\label{first}\\
\bar{N_i}=n_i.
\end{align}
\end{subequations}          
As in Poincare algebra we require that the rotation operator $\bar{M_i}$ leave time $\bar{X_0}$ invariant and rotate the space $\bar{X_i}$ as a vector. This means
\begin{subequations}\label{sha21}
\begin{align}
\left[\bar{M_i},\bar{X_0}\right]=0,\label{first}\\
\left[\bar{M_i},\bar{X_j}\right]=\epsilon_{ijk}\bar{X_k}.
\end{align}
\end{subequations}
Making use of the first equations of \eqref{sha20} and \eqref{sha3}, and \eqref{sha2}, we find that the space-time of the dual kappa Poincare is related to SR space-time via
\begin{subequations}\label{sha22}
\begin{align}
\bar{X_0}=\bar{f}(x_0,\vec{x}\cdot\vec{x}),\label{first}\\
\bar{X_i}=x_i\bar{g}(x_0,\vec{x}\cdot\vec{x}).
\end{align}
\end{subequations}
These are the counterpart of the relations given by \eqref{sha7} in kappa Poincare. The inverse of these relations are given by
\begin{subequations}\label{sha23}
\begin{align}
x_0=\bar{F}(\bar{X_0},\vec{\bar{X}}\cdot\vec{\bar{X}}),\label{first}\\
x_i=\bar{X_i}\bar{G}(\bar{X_0},\vec{\bar{X}}\cdot\vec{\bar{X}}),
\end{align}
\end{subequations}
which are the counterpart of the relations given by \eqref{sha10} in kappa Poincare. 

Substituting \eqref{sha22} in $\left[\bar{N_i},\bar{X_0}\right]$ and $\left[\bar{N_i},\bar{X_j}\right]$ and using \eqref{sha23}, the second of \eqref{sha3} and \eqref{sha2}, we find 
\begin{subequations}\label{sha24}
\begin{align}
\left[\bar{N_i},\bar{X_0}\right]=i\bar{D}(\bar{X_0},\vec{\bar{X}}\cdot\vec{\bar{X}})\bar{X_i},\label{first}\\
\left[\bar{N_i},\bar{X_j}\right]=i\delta_{ij}\bar{A}(\bar{X_0},\vec{\bar{X}}\cdot\vec{\bar{X}})+i\bar{X_i}\bar{X_j}\bar{B}(\bar{X_0},\bar{\vec{X}}\cdot\bar{\vec{X}}),
\end{align}
\end{subequations}
where
\begin{subequations}\label{sha25}
\begin{align}
\bar{A}=\bar{F}\bar{g}(\bar{x_0},\vec{x}\cdot\vec{x}),\label{first}\\
\bar{B}=\bar{G}^2\left[2\bar{F}\frac{\partial{\bar{g}}}{\partial{(\vec{x}\cdot\vec{x})}}+\frac{\partial{\bar{g}}}{\partial{x_0}}\right],\label{second}\\
\bar{D}=\bar{G}\left[2\bar{F}\frac{\partial{\bar{f}}}{\partial{(\vec{x}\cdot\vec{x})}}+\frac{\partial{\bar{f}}}{\partial{\bar{x_0}}}\right].
\end{align}
\end{subequations}
In the above equations, we have to make use of \eqref{sha23} to get the $\bar{X_0}$ and $\bar{\vec{X}}\cdot\bar{\vec{X}}$ dependence of $\bar{A}$, $\bar{B}$ and $\bar{C}$.       

Equations \eqref{sha24} and \eqref{sha25} in the dual kappa Poincare are the counterparts of equations \eqref{sha8} and \eqref{sha11} in kappa Poincare. Following a similar procedure in kappa Poincare, this time considering the Jacobi identity involving $\bar{N_i}$, $\bar{N_j}$ and $\bar{X_k}$, we find that $\bar{A}$, $\bar{B}$ and $\bar{C}$ satisfy
\begin{equation}\label{sha26}
\frac{\partial{\bar{A}}}{\partial{\bar{X_0}}}\bar{D}+2\frac{\partial{\bar{A}}}{\partial{(\vec{\bar{X}}\cdot\vec{\bar{X}})}}\left[\bar{A}+(\vec{\bar{X}}\cdot\vec{\bar{X}})\bar{B}\right]-\bar{A}\bar{B}=1.
\end{equation}
This equation is the counterpart of equation \eqref{sha9} in kappa Poincare theory. And just like in kappa Poincare, the fact that there is one (non-linear) equation for two unknown functions ($\bar{f}$ and $\bar{g}$ or their inverses $\bar{F}$ and $\bar{G}$) means that there is no unique dual kappa Poincare theory.

To complete the dual kappa Poincare algebra, we need to derive the dual phase space algebra, the action of rotations and boosts on the momenta, how the dual kappa Poincare algebra reduces to the Poincare limit (limiting value of the deformation parameter $\bar{\kappa}$).

This section will mirror the presentation in the latter part of Section II. By construction the dual kappa Poincare coordinates commute (see \eqref{sha22}), i.e., 
\begin{equation}\label{sha27}
\left[\bar{X_\mu},\bar{X_\nu}\right]=0.
\end{equation}
Since the space-time commutes, it follows that the momenta cocommutes, i.e.,
\begin{equation}\label{sha28}
\Delta(\bar{P_\mu})=\textbf{1}\otimes\bar{P_\mu}+\bar{P_\mu}\otimes\textbf{1}.
\end{equation}
This is the analogue of \eqref{sha12}. We define a non-trivial space-time coproduct given by
\begin{subequations}\label{sha29}
\begin{align}
\Delta(\bar{X_0})=\bar{X_0}\otimes\textbf{1}+\textbf{1}\otimes\bar{X_0},\label{first}\\
\Delta(\bar{X_i})=\bar{X_i}\otimes\textbf{1}+\exp{(-\bar{\kappa}\bar{X_0})}\otimes\bar{X_i}.
\end{align}
\end{subequations}
Comparing this with \eqref{sha13}, we see that this particular deformation of Poincare algebra can be appropriately called dual bicrossproduct basis. The parameter $\bar{\kappa}$ must have dimension of momenta and thus dual to the parameter $\kappa$ of the bicrossproduct basis, which has dimension of length. 

The Heisenberg double procedure yields the following phase space algebra
\begin{subequations}\label{sha30}
\begin{align}
\left[\bar{P_0},\bar{X_0}\right]=i,\label{first}\\
\left[\bar{P_i},\bar{X_j}\right]=-i\delta{ij},\label{second}\\
\left[\bar{P_i},\bar{X_0}\right]=\left[\bar{P_i},\bar{P_j}\right]=0,\label{third}\\
\left[\bar{P_0},\bar{X_i}\right]=-i\bar{\kappa}\bar{X_i},\label{fourth}\\
\left[\bar{P_0},\bar{P_i}\right]=i\bar{\kappa}\bar{P_i}.
\end{align}
\end{subequations}
From above, we see that in the dual bicrossproduct basis of the dual kappa Poincare algebra the energy does not commute with the momenta and spatial coordinates.

At this stage, we develop the dual bicrossproduct basis further by giving the counterpart formulas for \eqref{sha15} 
\begin{subequations}\label{sha31}
\begin{align}
\bar{f}=\frac{1}{\bar{\kappa}}\ln{(\bar{\kappa}x_0+\sqrt{\bar{\kappa}^2(x_0^2-\vec{x}\cdot\vec{x})+1})},\label{first}\\
\bar{g}=\dfrac{1}{\bar{\kappa}x_0+\sqrt{\bar{\kappa}^2(x_0^2-\vec{x}\cdot\vec{x})+1}},\label{second}\\
\bar{F}=\frac{1}{\kappa}\sinh{(\bar{\kappa}\bar{X_0})}+\frac{\bar{\kappa}}{2}\vec{\bar{X}}\cdot\vec{\bar{X}}\exp{(\bar{\kappa}\bar{X_0})},\label{third}\\
\bar{G}=\exp{(\bar{\kappa}\bar{X_0})}.
\end{align}
\end{subequations}
The main difference between \eqref{sha15} of the bicrossproduct basis and the above equations for the dual bicrossproduct is that $p_0^2-\vec{p}\cdot\vec{p}=m^2$ in the former while $x_0^2-\vec{x}\cdot\vec{x}$ does not have a corresponding invariant value in the latter and thus appears explicitly in the transformations. Using these expressions in \eqref{sha25}, we find the algebra of the boost generators with space-time in the dual bicrossproduct is given by \eqref{sha24} with
\begin{subequations}\label{sha32}
\begin{align}
\bar{A}=\frac{1}{2\bar{\kappa}}(1-\exp{(-2\bar{\kappa}\bar{X_0})})+\frac{\bar{\kappa}}{2}(\vec{\bar{X}}\cdot\vec{\bar{X}}),\label{first}\\
\bar{B}=-\bar{\kappa},\label{second}\\
\bar{D}=1.
\end{align}
\end{subequations}
The Poincare limit of the dual bicrossproduct basis of the dual kappa Poincare algebra is arrived at when $\bar{\kappa}\rightarrow 0$. In this limit we see that $\bar{A}\rightarrow \bar{X_0}$, $\bar{B}\rightarrow 0$. Clearly, the action of boosts in the limit $\bar{\kappa}\rightarrow 0$ gives back the Poincare limit (see the third and fourth equations of \eqref{sha4}).    

To complete the dual kappa Poincare algebra, we derive the action of boosts and rotations on the momenta. This is the analogue of equations \eqref{sha19}, which give the action of boosts and rotations on space-time in the bicrossproduct basis of kappa Poincare algebra. In this case, we have to derive the momenta in terms of the SR variables, which is the analogue of \eqref{sha17}. This involves finding $\bar{P_\mu}(x,p)$ that satisfy \eqref{sha30}. After a relatively long computation, we find
\begin{subequations}\label{sha33}
\begin{align}
\bar{P_0}=p_0\sqrt{\bar{\kappa}^2(x_0^2-\vec{x}\cdot\vec{x})+1},\label{first}\\
\bar{P_i}=p_i\sqrt{\bar{\kappa}^2(x_0^2-\vec{x}\cdot\vec{x})+1}-\bar{\kappa}(x_ip_0-x_0p_i).
\end{align}
\end{subequations}
Note, the dependence on space-time and momenta, which makes the momenta of the dual kappa Poincare non-commuting. Also, the transformation cannot be a canonical transformation because the fundamental brackets are not invariant. 

The inverse relations, the SR phase space variables in terms of the phase space variables of the dual bicrossproduct basis of the dual kappa Poincare algebra are given by
\begin{subequations}\label{sha34}
\begin{align}
p_0=\bar{P_0}\left[\cosh{(\bar{\kappa}\bar{X_0})}-\frac{\bar{\kappa}^2}{2}(\vec{\bar{X}}\cdot\vec{\bar{X}})\exp{(\bar{\kappa}\bar{X_0})}\right]^{-1},\label{first}\\
p_i=\exp{(-\bar{\kappa}\bar{X_0})}\bar{P_i}+\bar{\kappa}\bar{X_i}\bar{P_0}\left[\cosh{(\bar{\kappa}\bar{X_0})}-\frac{\bar{\kappa}^2}{2}(\vec{\bar{X}}\cdot\vec{\bar{X}})\exp{(\bar{\kappa}\bar{X_0})}\right]^{-1}.
\end{align}
\end{subequations}
The above equations give the rotation generators in the dual kappa Poincare algebra
\begin{equation}\label{sha35}
\bar{M_i}=\epsilon_{ijk}\bar{X_j}\bar{P_k}.
\end{equation}
Using this and the phase space algebra given by \eqref{sha30}, we find the action of rotations on the momenta as
\begin{subequations}\label{sha36}
\begin{align}
\left[\bar{M_i},\bar{P_0}\right]=0,\label{first}\\
\left[\bar{M_i},\bar{P_j}\right]=\epsilon_{ijk}\bar{P_k}.
\end{align}
\end{subequations}
Note, the action of rotations on the momenta is the same as in SR and DSR.

As for the action of boosts on the momenta, we will do the computations in SR phase space, make use of $\bar{N_i}=n_i$ and \eqref{sha33} and then transform back the results in the new kappa Poincare phase space variables. The results of the computations yield
\begin{subequations}\label{sha37}
\begin{align}
\left[\bar{N_i},\bar{P_0}\right]=i\bar{P_i}+i\bar{\kappa}\bar{N_i},\label{first}\\
\left[\bar{N_i},\bar{P_j}\right]=i\delta_{ij}\bar{P_0}+i\bar{\kappa}\epsilon_{ijk}\bar{M_k}.
\end{align}
\end{subequations}
Two comments are in order at this point. First, the action of boosts and rotations on the momenta in the dual bicrossproduct basis of the new kappa Poincare have the same structure as the action of boosts and rotations on the coordinates in the bicrossproduct basis of kappa Poincare (see \eqref{sha19}). Second, in the Poincare limit, i.e., $\bar{\kappa}\rightarrow 0$, we get back the action of boosts and rotations in SR.

To summarize, the dual kappa Poincare algebra, which we derived by making use of the dual bicrossproduct basis are given by the Lorentz algebra, \eqref{sha36} and \eqref{sha37}. The co-algebra are given by \eqref{sha28} and \eqref{sha29}. The phase space algebra is given by \eqref{sha30}. The action of rotations on the coordinates are given by \eqref{sha21} and the action of boosts on the coordinates are given by \eqref{sha24}. In the particular dual bicrossproduct basis, the right hand side of \eqref{sha24} are given by \eqref{sha32}.

\section{\label{sec:level4}Conclusion}
	In this paper , we have presented the dual kappa Poincare algebra. The advantage of this new way of modifying the Poincare algebra is that it has a space-time that is commuting but not co-commuting and a mometum space that is co-commuting but not commuting. Because the momenta has a trivial co-algebra, it will not suffer from the soccer ball and non-symmetric two particle momentum state problems. Since space-time commutes, there is no problem in defining the instantaneous velocity without resorting to various procedures in phase space \cite{Lukierski3}. Furthermore, if the analogy with Kowalski-Glikman's work, where he showed that the momentum space of DSR theories is De Sitter, holds, we may just have the space-time of this dual kappa Poincare algebra being De Sitter as the co-algebra given by \eqref{sha29} suggests. The parameter $\bar{\kappa}$ must be related to the cosmological constant $\Lambda$ through $\bar{\kappa}=\Lambda^{\frac{1}{2}}$ because for De Sitter space-time to be a solution of Einstein's equation $\Lambda=\frac{3}{l^2}$, where l is the De Sitter length \cite{Aldrovandi}. Note, the Poincare limit is achieved as $\Lambda\rightarrow0$. The other observer-independent scale to make the dual Poincare algebra a basis for a dual DSR is currently under investigation (a natural guess is the De Sitter length l) along with the co-algebra of boosts and rotations, the antipodes and co-units to have a complete Hopf algebra.   

\begin{acknowledgments}
The author would like to thank the Natural Science Research Institute and the Research and Creative Work Grant of
the University of the Philippines for supporting his research. I would like to thank the referee for his comments that led to the more appropriate classification of the algebra developed in this work.
\end{acknowledgments}

\end{document}